\documentclass[12pt]{article}
\usepackage{graphicx}
\def\bra#1{\left\langle #1\right|}
\def\ket#1{\left| #1\right\rangle}
\newcommand{\bers}{\begin{eqnarray*}}
\newcommand{\eers}{\end{eqnarray*}}
\newcommand{\bt}{\begin{itemize}}
\newcommand{\et}{\end{itemize}}
\def\beq{\begin{equation}}
\def\eeq{\end{equation}}
\def\bea{\begin{eqnarray}}
\def\eea{\end{eqnarray}}
\def\nn{\nonumber}
\def\sla#1{\raise.15ex\hbox{$/$}\kern-.57em #1}% Feynman slash

\def\sss{\scriptscriptstyle}

\def\bd{B_d^0}
\def\bs{B_s}
\def\bdbar{{\overline{B_d^0}}}
\def\bs{B_s^0}

\def\barp{{\raise.35ex\hbox
{${\sss (}$}}---{\raise.35ex\hbox{${\sss )}$}}}
\def\bdbarp{\hbox{$B_d$\kern-1.4em\raise1.4ex\hbox{\barp}}}
\def\bsbarp{\hbox{$B_s$\kern-1.4em\raise1.4ex\hbox{\barp}}}
\def\ks{K_{\sss S}}

\def\roughly#1{\mathrel{\raise.3ex\hbox
{$#1$\kern-.75em\lower1ex\hbox{$\sim$}}}}

\def\Act{{\cal A}_{ct}}
\def\Aut{{\cal A}_{ut}}

\def\aR{a_{\sss R}}
\def\epjc#1#2#3{{ Eur.\ Phys.\ J.}\ {\bf C#1}, #3 (#2)}

\def\npb#1#2#3{{ Nucl.\ Phys.} {\bf B#1}, #3 (#2)}
\def\plb#1#2#3{{ Phys.\ Lett.} {\bf #1B}, #3 (#2)}
\def\prd#1#2#3{{ Phys.\ Rev.} {\bf D#1}, #3 (#2)}
\def\newprd#1#2#3{{ Phys.\ Rev.} {\bf D#1}, #3 (#2)}
\def\prl#1#2#3{{ Phys.\ Rev.\ Lett.} {\bf #1}, #3 (#2)}

\def\zpc#1#2#3{{ Zeit.\ Phys.} {\bf C#1}, #3 (#2)}

\begin{document}

\begin{flushright}  
UdeM-GPP-TH-03-113 \\
McGill 03/22 \\
\end{flushright}

\begin{center}
\bigskip
{\Large \bf Extracting $\gamma$ from $\bd(t) \to D^{(*)+} D^{(*)-}$
and \vskip2truemm $\bd \to D_s^{(*)+} D^{(*)-}$ Decays} \\
\bigskip
\bigskip
{\large Alakabha Datta $^{a,}$\footnote{datta@physics.utoronto.ca} and
David London $^{b,c,}$\footnote{london@lps.umontreal.ca}}
\end{center}

\begin{flushleft}
~~~~~~~~~~~$a$: {\it Department of Physics, University of Toronto,}\\
~~~~~~~~~~~~~~~{\it 60 St.\ George Street, Toronto, ON, Canada M5S 1A7}\\
~~~~~~~~~~~$b$: {\it Physics Department, McGill University,}\\
~~~~~~~~~~~~~~~{\it 3600 University St., Montr\'eal QC, Canada H3A 2T8}\\
~~~~~~~~~~~$c$: {\it Laboratoire Ren\'e J.-A. L\'evesque, 
Universit\'e de Montr\'eal,}\\
~~~~~~~~~~~~~~~{\it C.P. 6128, succ. centre-ville, Montr\'eal, QC,
Canada H3C 3J7}
\end{flushleft}

\begin{center} 
\bigskip (\today)
\vskip0.5cm
{\Large Abstract\\}
\vskip3truemm
\parbox[t]{\textwidth} {We show that the CP phase $\gamma$ can be
obtained from measurements of $\bd(t) \to D^{(*)+} D^{(*)-}$ and $\bd
\to D_s^{(*)+} D^{(*)-}$. These decays are related by flavor SU(3) in
the limit where spectator-quark contributions are small. If the
pseudoscalar-pseudoscalar decays $\bd(t) \to D^+ D^-$ and $\bd \to
D_s^+ D^-$ are used, the leading-order SU(3)-breaking effect is
$f_{D_s}/f_D$. The dependence on decay constants can be removed by
using a double ratio involving two helicity states of the
vector-vector decays $\bd \to D^{*+} D^{*-}$ and $\bd \to D_s^{*+}
D^{*-}$. In this case the theoretical error arising from all sources
is in the range 5--10\%.}
\end{center}

\thispagestyle{empty}
\newpage
\setcounter{page}{1}
% Decrease texheight (for preprint numbers) again
%\textheight 23.0 true cm
\baselineskip=14pt

A great deal of work, both theoretical and experimental, has gone into
the study of CP-violating effects in the $B$ system. Within the
standard model (SM), CP violation is due to a complex phase in the
Cabibbo-Kobayashi-Maskawa (CKM) matrix, and this information is
elegantly encoded in the so-called unitarity triangle \cite{pdg}.
Measurements of CP violation in $B$ decays will allow us to extract
$\alpha$, $\beta$ and $\gamma$, the three interior angles of the
unitarity triangle \cite{CPreview}. By comparing the values of these
CP angles with the SM predictions, we will be able to test for the
presence of physics beyond the SM.

In general, weak phase information can only be extracted cleanly,
i.e.\ with no hadronic uncertainties, from modes which are dominated
by a single decay amplitude. One example is the so-called
``gold-plated'' mode $\bd(t)\to J/\psi \ks$, which is used to obtain
$\beta$. Other decays, such as $\bd(t)\to \pi^+ \pi^-$, receive both
tree and penguin contributions. In this case, more complicated
analyses, such as isospin \cite{isospin}, are necessary to cleanly
extract information about the CP phases \footnote{Note that, in fact,
there are both tree and penguin contributions to $\bd(t)\to J/\psi
\ks$ as well. However, in the Wolfenstein parameterization of the CKM
matrix \cite{Wolfenstein}, the weak phases of these two amplitudes are
equal. Thus, there is effectively only a single weak amplitude
contributing to $\bd\to J/\psi \ks$, and the extraction of $\beta$
from this decay mode is extremely clean.}.

Another decay mode which receives contributions from both tree and
penguin amplitudes is $\bd \to D^+ D^-$. If there were no penguin
contribution, the CP asymmetry in $\bd(t) \to D^+ D^-$ could be used
to obtain $\beta$, just like $\bd(t)\to J/\psi \ks$. However, the
penguin amplitude to this decay may well be important. Since it has a
different weak phase than the tree amplitude, the extracted value of
$\beta$ will not be the true value --- there will be some ``penguin
pollution.''

It is also possible to use decays in which one or both of the
final-state $D$'s is a vector meson. (If the final state $D^{*+}
D^{*-}$ is used, an angular analysis must be performed to separate the
different helicity states. We discuss this in more detail below.) The
BaBar experiment has measured the CP asymmetry in $\bd(t) \to D^{*+}
D^{*-}$ and finds $\sin 2\beta = 0.05 \pm 0.29~(stat) \pm 0.10~(syst)$
\cite{BaBar1}. This is to be compared with $\sin 2\beta = -0.73$
\cite{pdg}, the value expected based on measurements of CP violation
in $\bd(t)\to J/\psi \ks$. There is a deviation of about $2.5\sigma$,
suggesting that penguin pollution may well be important in the decay
$\bd(t) \to D^{(*)+} D^{(*)-}$. (BaBar has also measured CP violation
in the final state $D^{*\pm} D^\mp$ \cite{BaBar2}, but the errors are
still very large.)

Over the years, the decay $\bd \to D^{(*)+} D^{(*)-}$ has been
examined in some detail \cite{BDDbarrefs}. By studying the hadronic
properties of this decay, it is hoped that one can get some CP phase
information from its measurement. Most recently, it was pointed out
that, by comparing $\bd(t) \to D^+ D^-$ with its U-spin counterpart
$\bs \to D_s^+ D_s^-$, the phase $\gamma$ can be obtained
\cite{Fleischer}.

In this paper, we show that the weak phase $\gamma$ can be obtained by
comparing $\bd(t) \to D^+ D^-$ with $\bd \to D_s^+ D^-$, assuming that
$\beta$ is given by the CP asymmetry in $\bd(t)\to J/\psi \ks$. The
method itself is quite straightforward; the most important question is
the size of the theoretical uncertainty. The main point is that,
although $\bd \to D^+ D^-$ and $\bd \to D_s^+ D^-$ are technically not
related by flavor SU(3), the effective Hamiltonians describing them
are. Thus, to the extent that the contributions involving the
spectator quark are small, these decays {\it are} related by U-spin.
As a consequence, the main theoretical uncertainty --- SU(3) breaking
--- comes from the difference between the $D$ and $D_s$ decay
constants, which can in principle be measured or calculated on the
lattice. This theoretical error can be reduced further by using the
vector-vector modes $\bd(t) \to D^{*+} D^{*-}$ and $\bd \to D_s^{*+}
D^{*-}$ \cite{BKKbar}.

We begin by considering the pseudoscalar-pseudoscalar decay $\bd \to
D^+ D^-$. (The analysis applies equally to the case where one of the
final-state particles is a vector meson.) The amplitude for this decay
receives tree, exchange, $b\to d$ penguin and color-suppressed
electroweak penguin contributions \cite{su3}:
\bea
A^D & = & (T + E + P_c) \, V_{cb}^* V_{cd} + P_u \, V_{ub}^* V_{ud} +
(P_t + P_{EW}^C) \, V_{tb}^* V_{td} \nn\\
& = & (T + E + P_c - P_t - P_{EW}^C) \, V_{cb}^* V_{cd} + (P_u - P_t -
P_{EW}^C) \, V_{ub}^* V_{ud} \nn\\
& \equiv & \Act\ e^{i \delta^{ct}} + \Aut\ e^{i \gamma} e^{i
\delta^{ut}} ~,
\label{AmpDDbar}
\eea
where $\Act\equiv |(T + E + P_c - P_t - P_{EW}^C) V_{cb}^* V_{cd}|$,
$\Aut\equiv |(P_u - P_t - P_{EW}^C) V_{ub}^* V_{ud}|$, and we have
explicitly written out the strong phases $\delta^{ct}$ and
$\delta^{ut}$, as well as the weak phase $\gamma$. In the above, the
$P_i$ correspond to the $b\to d$ penguin amplitude with an internal
$i$-quark. The second line is obtained by using the unitarity of the
CKM matrix, $V_{ub}^* V_{ud} + V_{cb}^* V_{cd} + V_{tb}^* V_{td} = 0$,
to eliminate the $V_{tb}^* V_{td}$ term. The amplitude ${\bar A}^D$
for the decay $\bdbar \to D^+ D^-$ can be obtained from the above by
changing the signs of the weak phases. By making time-dependent
measurements of $\bd(t)\to D^+D^-$, one can obtain the three
observables
\bea
B &\equiv & \frac{1}{2} \left( |A^D|^2 + |{\bar A}^D|^2 \right) =
\Act^2 + \Aut^2 + 2 \Act \Aut \cos\delta \cos\gamma ~, \nn \\
a_{dir} &\equiv & \frac{1}{2} \left( |A^D|^2 - |{\bar A}^D|^2 \right)
= - 2 \Act \Aut \sin\delta \sin\gamma ~, \\
a_{indir} &\equiv & {\rm Im}\left( e^{-2i \beta} {A^D}^* {\bar A}^D
\right) = -\Act^2 \sin 2\beta - 2 \Act \Aut \cos\delta \sin (2 \beta +
\gamma) \nn\\
& & \hskip2.6truein
- \Aut^2 \sin (2\beta + 2 \gamma)~, \nn
\eea
where ${\delta}\equiv {\delta}^{ut} - {\delta}^{ct}$. It is
straightforward to count the number of theoretical parameters involved
in these experimental observables. There are five: the two magnitudes
$\Act$ and $\Aut$, one relative strong phase $\delta$, and two weak
phases ($\beta$ and $\gamma$). Even if we take $\beta$ from the CP
asymmetry in $\bd(t)\to J/\psi \ks$, there is still one more
theoretical parameter than there are observables. Thus, in order to
obtain weak phase information, it is necessary to add some theoretical
input \cite{CKMambiguity}.

This information can be obtained by considering the decay $\bd \to
D_s^+ D^-$. This decay receives tree, $b\to s$ penguin and
color-suppressed electroweak penguin contributions \cite{su3}:
\bea
A^{D_s} & = & (T' + P'_c) \, V_{cb}^* V_{cs} + P'_u \, V_{ub}^* V_{us}
+ (P'_t + P_{EW}^{\prime C})\, V_{tb}^* V_{ts} \nn\\
& = & (T' + P'_c - P'_t - P_{EW}^{\prime C}) \, V_{cb}^* V_{cs} +
(P'_u - P'_t - P_{EW}^{\prime C}) \, V_{ub}^* V_{us} \nn\\
& \approx & (T' + P'_c - P'_t - P_{EW}^{\prime C}) \, V_{cb}^* V_{cs}
\equiv \Act' e^{i \delta^{\prime ct}} ~.
\label{AmpDsDbar}
\eea
In the above, the $P'_i$ correspond to the $b\to s$ penguin amplitude
with an internal $i$-quark. The last line arises from the fact that
$V_{ub}^* V_{us}$ is much smaller than $V_{cb}^* V_{cs}$: $\left\vert
{V_{ub}^* V_{us} / V_{cb}^* V_{cs}} \right\vert \simeq 2\%$. Thus, the
measurement of the total rate for $\bd \to D_s^+ D^-$ yields $\Act'$
(the Particle Data Group gives $\Gamma(\bd \to D_s^+ D^-) = (8.0 \pm
3.0) \times 10^{-3}$ \cite{pdg}).

We now make the assumption that
\beq
\Delta \equiv {\sin\theta_c \, \Act' \over \Act} = {\sin\theta_c |(T'
+ P'_c - P'_t - P_{EW}^{\prime C}) V_{cb}^* V_{cs}| \over |(T + E +
P_c - P_t - P_{EW}^C) V_{cb}^* V_{cd}|} = 1 ~,
\label{assumption1}
\eeq
where $\sin\theta_c$ is the Cabibbo angle. Given that $\Act'$ is measured
in $\bd \to D_s^+ D^-$, $\Act$ can be obtained from the above
relation. This allows us to obtain $\gamma$ as follows. We first
introduce a fourth observable:
\bea
\aR & \equiv & {\rm Re}\left( e^{-2i \beta} {A^D}^* {\bar A}^D
\right) = \Act^2 \cos 2\beta + 2 \Act \Aut \cos\delta \cos (2 \beta +
\gamma) \nn\\
& & \hskip2.6truein
+ \Aut^2 \cos (2\beta + 2 \gamma)~.
\eea
The quantity $\aR$ is not independent of the other three observables:
\beq
\aR^2 = B^2 - a_{dir}^2 - a_{indir}^2 ~.
\eeq
Thus, one can obtain $\aR$ from measurements of $B$, $a_{dir}$ and
$a_{indir}$, up to a sign ambiguity. It is then straightforward to
obtain
\beq
\Act^2 = { \aR \cos(2\beta + 2\gamma) - a_{indir} \sin(2\beta +
2\gamma) - B \over \cos 2\gamma - 1} ~.
\eeq
Assuming that $2\beta$ is known from the measurement of CP violation
in $\bd(t)\to J/\psi \ks$, the assumption in Eq.~(\ref{assumption1})
therefore allows us to obtain $\gamma$.

What is the theoretical uncertainty inherent in such an assumption?
Referring to Eq.~(\ref{assumption1}), we note that the $E$ amplitude
in $\Act$ has been neglected, and that $\Act$ and $\Act'$ have been
related using flavor SU(3) symmetry. Now, strictly speaking, the
decays $\bd \to D^+ D^-$ and $\bd \to D_s^+ D^-$ are not related by
SU(3) --- the U-spin-transformed $\bd \to D^+ D^-$ decay is $\bs \to
D_s^+ D_s^-$ \cite{Fleischer}. However, the effective Hamiltonians
describing $\bd \to D^+ D^-$ and $\bd \to D_s^+ D^-$ {\it are} related
by SU(3). In this case, the spectator quark is not transformed by
U-spin between the two decays. Thus, contributions involving the
spectator quark are generally {\it not} related by
SU(3)\footnote{Relations between nonleptonic decays where the
spectator quark is not involved in the weak interaction were studied
in Ref.~\cite{dattalip}.}. One example of a process involving the
spectator quark is the exchange diagram, $E$, which contributes to
$\bd \to D^+ D^-$ but not to $\bd \to D_s^+ D^-$. As long as $E$ is
small, it is a good approximation to relate the two decays by SU(3).

We first estimate the size of the exchange contribution $E$ using
factorization:
\bea
E_{fac} & \sim & \frac{G_{\sss F}}{\sqrt{2}} \bra{0} \bar{d}
\gamma_{\mu}(1-\gamma_5) b \ket{\bd} \bra{D^+D^-} \bar{c}
\gamma^{\mu}(1-\gamma_5) c \ket{0} \nn\\
& \sim & f_B (P_{\sss D}^++P_{\sss D}^-)_{\mu} \bra{D^+D^-} \bar{c}
\gamma^{\mu} c \ket{0} \nn\\
& = & 0 ~,
\label{EF}
\eea
where we have used current conservation in the last line above. Thus,
the only contributions to $E$ come from nonfactorizable effects
\cite{fajfer}. However, these calculations are highly model-dependent
and no definite conclusions can be drawn from them\footnote{When one
or both of the final-state $D$'s is a vector meson, then $E_{fac}$
does not vanish. However, it is still suppressed by small Wilson
coefficients and form factors. There can also be contributions to $E$
from nonfactorizable effects \cite{fajfer}.}. Naive estimates put the
size of exchange contributions at about 5\% \cite{su3}. The most
prudent approach is to rely on experimental measurements. For example,
the decays $\bd \to D_s^+ D_s^-$ and $\bd \to D^0 {\bar D}^0$ proceed
mainly through weak interactions involving the spectator quark. Hence,
a measurement of these rates relative to that of $\bd \to D^+ D^-$
will provide an estimate of how important the $E$ contributions are.
Henceforth we will neglect these contributions: our dynamical
assumption is therefore that the decays $\bd \to D^+ D^-$ and $\bd \to
D_s^+ D^-$ are dominated by contributions in which the spectator quark
is not involved.

With this assumption, $\bd \to D^+ D^-$ and $\bd \to D_s^+ D^-$ are
related by SU(3), and Eq.~\ref{assumption1} holds in the SU(3) limit.
However, we know that SU(3) breaking is typically about 25\%, and the
important task at this stage is to examine the source(s) of such
breaking. Now, we know that $(P_c - P_t - P_{EW}^C)/T$ and $(P'_c -
P_t' - P_{EW}^{\prime C})/T'$ are both less than unity, probably also
$\sim 25\%$. Thus, SU(3) breaking originates mainly in the ratio
$T'/T$ in Eq.~\ref{assumption1} --- the SU(3)-breaking contribution
from the penguin amplitudes is of higher order.

We now turn to a calculation of the color-allowed tree amplitudes $T$
and $T'$. As a first step, we use factorization. In the SM, the tree
amplitudes for $\bd \to D_s^+D^{-}$ and $\bd \to D^{+}D^{-}$ are
generated by two terms in the effective Hamiltonian \cite{buras}:
\beq
H_{tree}^q = {G_F \over \sqrt{2}} \left[ V_{ub}V^*_{uq}(c_1 O_1^q + c_2
O_2^q) \right] ~,
\label{Heff}
\eeq
where $q$ can be either a $d$ quark ($\bd \to D^{+}D^{-}$) or an $s$
quark ($\bd \to D_s^+D^{-}$). The operators $O_i^q$ are defined as
\beq
O_1^q = {\bar q}_\alpha \gamma_\mu L u_\beta \, {\bar u}_\beta
\gamma^\mu L b_\alpha ~~,~~ O_2^q = {\bar q} \gamma_\mu L u \, {\bar
u} \gamma^\mu L b ~,
\eeq
where $R(L) = 1 \pm \gamma_5$. The values of the Wilson coefficients
$c_1$ and $c_2$ can be found in Ref.~\cite{buras}. 

Within factorization, the tree amplitude for $\bd \to D_s^+ D^-$ is
given by
\beq
{\cal M} = {G_F \over \sqrt{2}} V_{cs}^{*} V_{cb}
\left(\frac{c_1}{N_c} + c_2 \right) \langle D_s^+|\, \bar{s}
\gamma_\mu(1-\gamma_5) \, c\, |\, 0 \rangle \langle D^- |\, \bar{c} \,
\gamma^\mu (1-\gamma_5) \, b\, |\, \bd \rangle ~,
\label{amp}
\eeq
where $N_c$ represents the number of colors. The factorized tree
amplitude for $\bd \to D^+ D^-$ is as above, but with the $s$ quark
replaced by a $d$ quark. The currents in Eq.~(\ref{amp}) are given by
\bea
\langle D_s^+|\, \bar{s} \gamma_\mu(1-\gamma_5) \, c\, |\, 0 \rangle &
= & i f_{D_s} q_\mu ~, \nn\\
\langle D^-(p_D) |\, \bar{c} \, \gamma^\mu (1-\gamma_5) \, b\, |\,
\bd(p_B) \rangle & = & \left[ (p_B+ p_D)_\mu - \frac{m_B^2-m_D^2}{q^2}
q_\mu \right] F_1(q^2) \nn\\
& & \hskip0.5truein + \frac{m_B^2-m_D^2}{q^2} q_\mu F_0(q^2) ~,
\eea
where $q = p_B - p_D $, and $F_{0,1}$ are form factors \cite{BSW}.
For $B \to D$ transitions it is more appropriate to consider the form
factors as functions of $ \omega = (m_B^2 + m_D^2 -q^2)/(2 \, m_B \,
m_D)$.

The tree-level matrix elements for $\bd \to D_s^+ D^- $ and $\bd \to
D^+ D^- $ are then given by
\bea
|T'( \bd \to D_s^+ D^-)| & = & {\frac{G_F}{\sqrt 2}} |V_{cb}V_{cs}^*|
\left( {c_1 \over N_c} + c_2 \right) f_{D_s} F_0(\omega_s)
(m_B^2-m_D^2) ~, \nn\\
|T( \bd \to D^+ D^-)| & = & \frac{G_F}{\sqrt 2} |V_{cb}V_{cd}^*|
\left( {c_1 \over N_c} + c_2 \right) f_{D} {F_0(\omega_d)}
(m_B^2-m_D^2) ~,
\eea
where 
\beq
\omega_s = \frac{m_B^2+m_D^2-m_{D_s}^2}{2 \, m_B \, m_D} ~~,~~
\omega_d = \frac{m_B^2}{2 \, m_B \, m_D} ~.
\eeq
Thus, as far as SU(3) breaking is concerned, Eq.~(\ref{assumption1})
is given by
\beq
\Delta \simeq {\sin\theta_c |T' V_{cb}^* V_{cs}| \over |T V_{cb}^* V_{cd}|}
= \frac{F_0(\omega_s)}{F_0(\omega_d)} \, \frac{f_{D_s}}{f_D} ~,
\label{leadingSU3}
\eeq
with $\omega_d=1.42$ and $\omega_s=1.4$. The form factor $F_0$ can be
related to the Isgur-Wise function in the heavy quark limit. Since
$F_0$ is smooth with no sudden sharp changes, we can, with negligible
error, set $F_0(\omega_d)=F_0(\omega_s)$. The upshot is that, within
factorization, SU(3) breaking is due almost entirely to the difference
between the $D$ and $D_s$ decay constants. This ratio has been
measured quite precisely on the lattice: $f_{D_s}/f_D = 1.22 \pm 0.04$
\cite{lattice}. If one uses this value, the leading-order theoretical
error in Eq.~(\ref{assumption1}) is quite small. If one does not wish
to rely on such calculations, the ratio $f_{D_s}/f_D$ can in principle
be measured, although the error is likely to be considerably larger
than in the lattice calculation.

We therefore conclude that, within factorization, the principal
contribution to SU(3) breaking in Eq.~(\ref{assumption1}) is $\Delta
\simeq f_{D_s}/f_D$. The error in this conclusion is related to the
uncertainty in the factorization assumption. Now, factorization is
expected to reliably predict the color-allowed tree and penguin
amplitudes. However, the rescattered penguins $P_{u,c}$ and $P'_{u,c}$
are estimated within factorization using the perturbative piece only.
It is quite possible that additional nonperturbative contributions are
present. Nevertheless, our main argument still stands: even with such
nonperturbative contributions, the penguin amplitudes are still
smaller than the tree amplitude. That is, $(P_c - P_t - P_{EW}^C)/T$
and $(P'_c - P_t' - P_{EW}^{\prime C})/T'$ can still be treated as
small quantities, in which case the conclusion that $\Delta \simeq
f_{D_s}/f_D$ remains valid -- the SU(3)-breaking corrections from the
rescattered penguins are a second-order effect. If these penguin
contributions are large, i.e.\ $(P_c - P_t - P_{EW}^C) \sim T$ and/or
$(P'_c - P_t' - P_{EW}^{\prime C}) \sim T'$, one would expect to see
large direct CP violation and/or a large discrepancy in the
measurement of $\sin 2\beta$ in $\bd(t) \to D^+D^-$ and $\bd(t)\to
J/\psi \ks$.

Factorization has been used to study $B \to D^{(*)} {\bar D}^{(*)}$
decays \cite{luorosner}, and it has been found that experiments are
consistent with the factorization predictions. In addition,
Ref.~\cite{lukewise} discusses tests indicating that factorization in
$B \to D^{(*)} X$ is a consequence of large $N_c$ QCD.  These two
analyses suggest that factorization in $B \to D^{(*)} {\bar D}^{(*)}$
is a consequence of large $N_c$ QCD \cite{thooft,buraslargeN}, with
nonfactorizable corrections arising at $O(1/N_c^2)$. Therefore, in the
large $N_c$ approach to nonleptonic decays, the deviation from $\Delta
\simeq f_{D_s}/f_D$ in Eq.~(\ref{assumption1}) is suppressed by
$1/N_c^2$ as well as $SU(3)$ breaking\footnote{Note that the exchange
diagrams are suppressed by $1/N_c$ relative to the factorized
amplitude. However, as discussed earlier, these contributions are
suppressed by additional effects and can be ignored.}. In other words,
nonfactorizable effects lead to
\beq
\Delta = \frac{f_{D_s}}{f_D} + (a_s-a_d) \frac{1}{N_c^2} =
\frac{f_{D_s}}{f_D} + a\frac{m_s}{\Lambda_\chi}\frac{1}{N_c^2} \simeq
\frac{f_{D_s}}{f_D} \left( 1 + a\frac{m_s}{\Lambda_\chi}\frac{1}{N_c^2}
\right) ~,
\label{error}
\eeq
where $a \sim O(1)$ and $\Lambda_\chi$ is the chiral symmetry breaking
scale. In the $N_c \to \infty$ limit, factorization holds and so
$\Delta=f_{D_s}/f_D$. Furthermore, in the $SU(3)$ limit the $1/N_c^2$
corrections to $\Delta-1$ vanish. We therefore expect {\it all}
nonfactorizable corrections to $\Delta \simeq f_{D_s}/f_D$ to be quite
small, around 3\%.

One of the key ingredients in this method is the claim that the
leading-order SU(3) breaking is given by $f_{D_s}/f_D - 1$, which is
about 25\%. As argued above, this follows from factorization, which
has been found to hold in $B \to D^{(*)} {\bar D}^{(*)}$ decays.
However, one might worry that this estimate of SU(3) breaking is too
small, based on what happens in superficially similar $D$ decays. For
example, using the above arguments, we would expect that
$\sin^2\theta_c \, BR(D^0 \to K^- \pi^+) / BR(D^0 \to K^- K^+) \simeq
(f_\pi/f_K)^2 = 0.67$. But experimentally this ratio is found to take
the value $0.45 \pm 0.02$ \cite{pdg}, showing that SU(3) breaking here
is about 50\%.  The problem with this reasoning is that there are
significant differences between $D$ and $B$ decays. In particular,
factorization is badly broken in $D^0 \to K^-\pi^+/K^-K^+$, where
large rescattering effects are present, perhaps from nearby resonances
\cite{Dnonlep}.  In addition, exchange contributions, which are higher
order in $1/N_c$, are significant in many $D$ decays \cite{LipClose},
again possibly from nearby resonance effects. On the other hand, at
the $B$ mass, which is far above the resonance region, there is no
evidence of large rescattering or of large exchange diagrams
\cite{luorosner}.  Thus, there are sizeable effects in $D$ decays, not
present in $B$ decays, from nonfactorizable contributions and from
exchange diagrams. In general, these will lead to larger SU(3)
breaking. For this reason, $D$ decays do not furnish a reliable
estimate of SU(3)-breaking effects in $B$ decays.

To summarize, the CP phase $\gamma$ can be extracted from a study of
the decays $\bd \to D^+ D^-$ and $\bd \to D_s^+ D^-$. We neglect the
exchange contribution to $A(\bd \to D^+ D^-)$ and the $V_{ub}^*
V_{us}$ piece of $A(\bd \to D_s^+ D^-)$. The main contribution to the
theoretical error comes from SU(3) breaking in the ratio of these two
amplitudes. To leading order, it is given by $f_{D_s}/f_D$
[Eq.~(\ref{error})]. Higher-order SU(3)-breaking corrections come from
penguin amplitudes and nonfactorizable contributions. If $f_{D_s}/f_D$
is known precisely, the overall theoretical error is in the range
5--10\%.

It is possible to eliminate completely the dependence on decay
constants by considering vector-vector decays. For decays such as $\bd
\to D^{*+} D^{*-}$, there are three helicity states. These helicity
amplitudes follow simply from Eq.~(\ref{AmpDDbar}):
\beq
A^{D^*}_\lambda = \Act^\lambda \, e^{i \delta^{ct}_\lambda} +
\Aut^\lambda \, e^{i \gamma} e^{i \delta^{ut}_\lambda} ~,
\label{AmpD*}
\eeq
where the helicity index $\lambda$ takes the values $\left\{
0,\|,\perp \right\}$. Using CPT invariance, the full decay amplitudes
can be written as
\bea
{\cal A}^{D^*} &=& A_0^{D^*} g_0 + A_\|^{D^*} g_\| + i \,
A_\perp^{D^*} g_\perp~, \nn\\
{\bar{\cal A}}^{D^*} &=& {\bar A}_0^{D^*} g_0 + {\bar A}_\|^{D^*} g_\|
- i \, {\bar A}_\perp^{D^*} g_\perp~,
\label{fullamps}
\eea
where the $g_\lambda$ are the coefficients of the helicity amplitudes
written in the linear polarization basis. The $g_\lambda$ depend only
on the angles describing the kinematics \cite{glambda}. The
time-dependent decay rates can now be written as
\bea
\Gamma(\bd(t) \to D^{*+}D^{*-}) & = & e^{-\Gamma t}
\sum_{\lambda\leq\sigma} \Bigl(\Lambda_{\lambda\sigma} \pm
\Sigma_{\lambda\sigma}\cos(\Delta M t)\nn\\ 
&& \hskip1.0truein \mp \rho_{\lambda\sigma}\sin(\Delta M t)\Bigr)
g_\lambda g_\sigma ~.
\eea
By performing a time-dependent angular analysis of the decay $\bd(t)
\to D^{*+}D^{*-}$, one can measure the 18 observables
$\Lambda_{\lambda\sigma}$, $\Sigma_{\lambda\sigma}$ and
$\rho_{\lambda\sigma}$.

In fact, not all of these 18 observables are independent. There are a
total of six amplitudes describing the decays $\bd \to D^{*+} D^{*-}$
and $\bdbar \to D^{*+} D^{*-}$: $A^{D^*}_\lambda$ and ${\bar
A}^{D^*}_\lambda$. At best, one can measure the magnitudes and
relative phases of these six amplitudes, giving only 11 independent
measurements. On the other hand, one can see from Eq.~(\ref{AmpD*})
that these 11 observables are described by 13 theoretical parameters:
the six magnitudes $\Act^\lambda$ and $\Aut^\lambda$, five relative
strong phases, and the two weak phases $\beta$ and $\gamma$. As
before, even assuming that $\beta$ has been measured in $\bd(t)\to
J/\psi \ks$, there is still one more theoretical parameter than there
are measurements. Once again, it is necessary to add theoretical
input.

To obtain such input, we consider the decay $\bd \to D_s^{*+}
D^{*-}$. Analogous to Eq.~(\ref{AmpDsDbar}), the helicity amplitudes
can be written as
\beq
A^{D_s^*}_\lambda = \Act^{\prime\lambda} \, e^{i \delta^{\prime
ct}_\lambda} ~.
\label{AmpDs*}
\eeq
We define
\beq
\Delta_{\lambda} \equiv {\sin\theta_c \, \Act^{\prime\lambda} \over
\Act^\lambda} ~.
\eeq
The theoretical input is now provided by the assumption that
\beq
\Delta' \equiv \frac{ \Delta_{\lambda'}}{ \Delta_{\lambda}} = 1 ~.
\label{assumption2}
\eeq
The advantage of this assumption is that, because we are considering a
double ratio, much of the theoretical error cancels. For example, note
that $\Delta_\lambda$ and $\Delta_{\lambda'}$ are each analogous to
the quantity $\Delta$, defined for the case of $\bd(t) \to D^+ D^-$
and $\bd \to D_s^+ D^-$ [Eq.~(\ref{assumption1})]. Thus, to leading
order in SU(3) breaking, the decay constants cancel in the ratio
$\Delta'$ [see Eq.~(\ref{leadingSU3})], so that
Eq.~(\ref{assumption2}) holds.

Since the leading-order SU(3)-breaking effects cancel in $\Delta'$, we
now examine second-order effects. We can write
\beq
\Delta_{\lambda} = {\sin\theta_c \, \left\vert T'_\lambda + P'_\lambda
\right\vert \over \left\vert T_\lambda + P_\lambda \right\vert} =
\sin\theta_c {\left\vert T'_\lambda \right\vert \over \left\vert
T_\lambda \right\vert} {\sqrt{1 + {z'_\lambda}^2 + 2 z'_\lambda
\cos\Delta_s^\lambda} \over \sqrt{1 + {z_\lambda}^2 + 2 z_\lambda
\cos\Delta_d^\lambda}} ~.
\eeq
In the above, $z'_\lambda \equiv |P'_\lambda/T'_\lambda|$ and
$z_\lambda \equiv |P_\lambda/T_\lambda|$, where $P' \equiv (P'_c -
P_t' - P_{EW}^{\prime C})$ and $P \equiv (P_c - P_t - P_{EW}^C)$. As
mentioned earlier, we expect $z'_\lambda,~z_\lambda \sim 25\%$. Also,
$\Delta_s^\lambda$ ($\Delta_d^\lambda$) is the relative phase between
$P'_\lambda$ and $T'_\lambda$ ($P_\lambda$ and $T_\lambda$). Now,
$T'_\lambda$ and $T_\lambda$ are related by SU(3):
$T'_\lambda/T_\lambda = f_{D_s^*}/f_{D^*}$
[Eq.~(\ref{leadingSU3})]. Similarly, $z'_\lambda$ and $z_\lambda$ are
related by SU(3): we can write $z'_\lambda = z_\lambda (1 +
r_\lambda)$, where $r_\lambda$ is the SU(3)-breaking term. We expect
that $r_\lambda \sim 25\%$. Putting all the pieces together, we obtain
\beq
\Delta_\lambda \simeq \sin\theta_c {f_{D_s^*} \over f_{D^*}} \left[ 1 +
z_\lambda (\cos\Delta_s^\lambda - \cos\Delta_d^\lambda) + z_\lambda
r_\lambda \cos\Delta_s^\lambda \right].
\eeq
Thus,
\beq
\Delta' = {1 + z_\lambda (\cos\Delta_s^\lambda - \cos\Delta_d^\lambda)
+ z_\lambda r_\lambda \cos\Delta_s^\lambda \over 1 + z_{\lambda'}
(\cos\Delta_s^{\lambda'} - \cos\Delta_d^{\lambda'}) + z_{\lambda'}
r_{\lambda'} \cos\Delta_s^{\lambda'}} ~.
\eeq
{}From this expression, we see that if $z_{\lambda}$ and the SU(3)
corrections, $r_{\lambda}$, are helicity-independent, we have $\Delta'
= 1$ [Eq.~(\ref{assumption2})].  Even if the $z_{\lambda}$ and
$r_{\lambda}$ do depend on the helicity, it is possible that there
will be cancellations in $\Delta'$, though this is not guaranteed. The
most conservative thing to say is that the corrections from penguin
amplitudes to $\Delta' = 1$ are at the level of $|P/T|
(m_s/\Lambda_\chi) \sim 5\%$.

Finally, we consider nonfactorizable corrections to the leading-order
term. Following Eq.~(\ref{error}) we can write
\bea
\Delta_{\lambda} & = & \frac{f_{D_s^*}}{f_{D^*}} \left( 1 + \left(
a_{s,\lambda} - a_{d,\lambda} \right) {1 \over N_c^2} \right) =
\frac{f_{D_s^*}}{f_{D^*}} \left( 1 + a_\lambda
\frac{m_s}{\Lambda_\chi}\frac{1}{N_c^2} \right) ~, \nn\\
\Delta_{\lambda'} & = & \frac{f_{D_s^*}}{f_{D^*}} \left( 1 + \left(
a_{s,\lambda'} - a_{d,\lambda'} \right) {1 \over N_c^2} \right) =
\frac{f_{D_s^*}}{f_{D^*}} \left( 1 + a_{\lambda'}
\frac{m_s}{\Lambda_\chi}\frac{1}{N_c^2} \right) ~,
\label{deltalambda}
\eea
where $a_{s,\lambda}$ and $a_{d,\lambda}$ are the nonfactorizable
$1/N_c^2$ corrections to the {\it tree} amplitude in $\bd \to D_s^{*+}
D^{*-}$ and $\bd \to D^{*+} D^{*-}$ for the helicity $\lambda$, and
similarly for $a_{s,\lambda'}$ and $a_{d,\lambda'}$. We therefore
obtain
\bea
\Delta' = 1 + (a_{\lambda'} - a_{\lambda})
\frac{m_s}{\Lambda_\chi}\frac{1}{N_c^2} ~.
\eea
This shows that if one has nonfactorizable $1/N_c^2$ corrections that
are independent of the helicity states, $\Delta' = 1$ even with
second-order SU(3) breaking.  For nonfactorizable $1/N_c^2$
corrections that are different for different helicity states, nothing
definite can be said about the signs and magnitudes of $a_{\lambda}$
and $a_{\lambda'}$. In this case, we expect nonfactorizable
corrections to the leading term at the level of 3\% or less.

There are therefore four sources of theoretical error in
Eq.~(\ref{assumption2}): the neglect of the exchange contribution in
$\bd \to D^{*+}D^{*-}$, the neglect of the $|V_{ub}^*V_{us}|$ term in
$\bd \to D_s^{*+}D^{*-}$, SU(3)-breaking in the penguin corrections,
and nonfactorizable corrections. All errors are small, and we expect a
net violation of the relation $\Delta' = 1$ at the level of 5--10\%.

In conclusion, we have shown that $\gamma$ can be obtained from the
time-dependent measurement of the decay $\bd(t) \to D^{(*)+}
D^{(*)-}$, along with the rate for $\bd \to D_s^{(*)+} D^{(*)-}$. We
have assumed that $\beta$ has been measured in $\bd(t)\to J/\psi \ks$.
The method relies on the fact that $\bd \to D^{(*)+} D^{(*)-}$ and
$\bd \to D_s^{(*)+} D^{(*)-}$ are related by flavor SU(3) in the limit
where exchange contributions are negligible. If one uses the
pseudoscalar-pseudoscalar decays $\bd(t) \to D^+ D^-$ and $\bd \to
D_s^+ D^-$, the leading-order SU(3)-breaking effect is simply the
ratio of decay constants $f_{D_s}/f_D$. The value for this ratio
can be taken from lattice calculations (with a tiny error), in which
case the overall theoretical uncertainty in this method is around
5--10\%. Alternatively, the decay constants can in principle be
measured. In this case the accuracy of the method is limited by the
precision on the measurements of the decay constants.

One can reduce the theoretical error by using vector-vector final
states and performing an angular analysis to measure two different
helicity states. The method then uses a double ratio of $\bd \to
D^{*+} D^{*-}$ and $\bd \to D_s^{*+} D^{*-}$ measurements. Because we
consider a double ratio, all dependence on the ratio
$f_{D_s^*}/f_{D^*}$ cancels. Further SU(3) corrections are suppressed,
either by $P/T$ ratios, or by $1/N_c^2$. Although we expect that there
will be some cancellation in the SU(3)-breaking effects in the double
ratio, this is a model-dependent conclusion. Conservatively, the
theoretical error in this method is 5--10\%.

\bigskip
\noindent
{\bf Acknowledgements}:
%\bigskip
We thank Timothy Gershon for asking the questions which led to this
study. A.D. thanks D.L. for the hospitality of the Universit\'e de
Montr\'eal, where part of this work was done. The work of D.L. was
financially supported by NSERC of Canada.

%%%%%%%%%%%%%%%%%%%%% REFERENCES %%%%%%%%%%%%%%%%%%%%%%%%%%%%%%%%

\end{document}